\documentclass[nofootinbib,twocolumn,showpacs,preprintnumbers,amsmath,amssymb]{revtex4}
\usepackage{amsmath,amssymb,graphicx}
\usepackage{graphicx}
\usepackage{dcolumn}
\usepackage{bm}
\usepackage{epsfig}
\usepackage{here}
\usepackage{float}
\usepackage{amsmath}
\usepackage[usenames]{color}

\newcommand{\bea}{\begin{eqnarray}}
\newcommand{\eea}{\end{eqnarray}}

\begin{document}
\draft

\title{Newtonian hydrodynamic equations with relativistic pressure and velocity}
\author{Jai-chan Hwang${}^{1}$, Hyerim Noh${}^{2}$, J\'ulio Fabris${}^{3,4}$, Oliver F. Piattella${}^{3}$, Winfried Zimdahl${}^{3}$ }
\address{${}^{1}$Department of Astronomy and Atmospheric Sciences,
                 Kyungpook National University, Taegu, Korea \\
         ${}^{2}$Korea Astronomy and Space Science Institute,
                 Daejon, Korea \\
         ${}^{3}$Departamento de Fisica, Universidade Federal do Espirito Santo, Vit\'oria, Brasil \\
         ${}^{4}$National Research Nuclear University ¡°MEPhI¡±,
                 Kashirskoe sh. 31, Moscow 115409, Russia
         }


\begin{abstract}

We present a new approximation to include fully general relativistic pressure and velocity in Newtonian hydrodynamics. The energy conservation, momentum conservation and two Poisson's equations are consistently derived from Einstein's gravity in the zero-shear gauge assuming weak gravity and action-at-a-distance limit. The equations show proper special relativity limit in the absence of gravity. Our approximation is complementary to the post-Newtonian approximation and the equations are valid in fully nonlinear situations.

\end{abstract}


\maketitle

%
%

\vskip .1cm
{\bf 1. Introduction:}
Considering the enormous practical and conceptual difficulties in handling general relativistic astrophysical situations using numerical simulations of full Einstein's gravity \cite{numerical-relativity}, it is always welcome to have an approximation method. The post-Newtonian (PN) approximation is one such method \cite{Chandrasekhar-1965,Weinberg-1972,PN,HNP-2008-PN} where we restore the good and old absolute space and absolute time, and regard Einstein's gravity effects as corrections to the Newtonian equations. In this way we can handle weak but relativistic effects of gravity in Newtonian style, and the resulting equations are fully nonlinear.

Here, we provide a complementary approximation which can handle the fully relativistic pressure and velocity in the weak gravity and action-at-a-distance limit: for our assumptions see Eq.\ (\ref{assumptions}). We present the energy conservation, momentum conservation and Poisson's equation which allow us to handle such astrophysical situations in Newtonian manner. In this approximation also the equations are valid to fully nonlinear orders. Our derivation is based on the {\it zero-shear gauge} which will be explained later. We {\it ignore} the transverse-tracefree tensor-type perturbation in the spatial metric, and ignore the anisotropic stress.


%
%

\vskip .1cm
{\bf 2. Result:}
A closed form of new hydrodynamic equations we are proposing is
\bea
   & & {d \over dt} \varrho
       + \left( \varrho + {p \over c^2} \right)
       \nabla \cdot {\bf v}
       = {1 \over c^2} \left( {d p \over dt}
       - {1 \over \gamma^2} \dot {p} \right),
   \label{E-conservation} \\
   & & {d \over dt} {\bf v}
       = \nabla \Phi
       - {1 \over \varrho + p/c^2}
       {1 \over \gamma^2}
       \left( \nabla p
       + {1 \over c^2} {\bf v} \dot {p} \right),
   \label{M-conservation} \\
   & & \Delta \Phi
       + 4 \pi G \varrho
       = {12 \pi G \over c^2}
   \nonumber \\
   & & \qquad \times
       \Delta^{-1} \nabla_i \nabla^j
       \left[ \left( \varrho + {p \over c^2} \right)
       \gamma^2 \left( {v^i v_j}
       - {2 \over 3} \delta^i_j {v^2} \right) \right],
   \label{Poisson-ZSG}
\eea
with
\bea
   {d \over dt} \equiv {\partial \over \partial t}
       + {\bf v} \cdot \nabla, \quad
       \gamma = {1 \over \sqrt{ 1 - {v^2 / c^2}}},
   \label{gamma}
\eea
where $\varrho$, $p$ and ${\bf v}$ are the density, pressure and velocity, respectively, and $\gamma$ is the Lorentz factor with $v^2 \equiv {\bf v} \cdot {\bf v} \equiv v^i v_i$; $\Delta^{-1}$ is an inverse Laplacian operator, and an overdot indicates the partial time derivative with respect to $t$. In the absence of gravity Eqs.\ (\ref{E-conservation}) and (\ref{M-conservation}) properly reproduce the special relativistic hydrodynamics, see below Eq.\ (\ref{M-conservation-4}). Notice that the role of gravity is rather trivial in these special relativistic conservation equations, while the role of special relativity (especially the velocity) is nontrivial in the Poisson's equation, see Eq.\ (\ref{Poisson-ZSG}): for non-relativistic velocity we recover the ordinary Poisson's equation (without pressure correction!) known in Newtonian gravity.

Our metric convention is
\bea
   ds^2 = - \left( 1 - {2 \Phi \over c^2} \right) c^2 d t^2
       + \left( 1 + {2 \Psi \over c^2} \right) \delta_{ij} d x^i d x^j,
   \label{metric}
\eea
where $\Phi$ and $\Psi$ are the Newtonian and the post-Newtonian gravitational potentials, respectively; $\Psi$ is determined by a separate Poisson-like equation
\bea
   & & \hskip -.5cm
       \Delta \Psi
       + 4 \pi G \varrho
       = - 4 \pi G \left(
       \varrho + {p \over c^2} \right)
       \left( \gamma^2 - 1 \right).
   \label{Poisson-Psi}
\eea
We may call Eqs.\ (\ref{Poisson-ZSG}) and (\ref{Poisson-Psi}) as the Newtonian and the post-Newtonian Poisson's equations, respectively. The relativistic velocity causes a difference between $\Phi$ and $\Psi$.

It is important to notice that the zero-shear gauge condition alone does {\it not} allow us to write the metric in this simple form: a proof of the case will be given after full calculation with the general metric, see Eq.\ (\ref{metric-2}), below Eq.\ (\ref{Phi-Psi-2}) and below Eq.\ (\ref{kappa-def}). In this work we will derive the above set of equations consistently from the full Einstein's equations.

The weak gravity and the action-at-a-distance {\it assumptions} are
\bea
   {\Phi \over c^2} \ll 1, \quad
       {\Psi \over c^2} \ll 1, \quad
       \gamma^2 {t_\ell^2 \over t_g^2} \ll 1,
   \label{assumptions}
\eea
where $t_g \sim 1/\sqrt{G \varrho}$ and $t_\ell \sim \ell/c \sim 1/(kc)$ are the gravitational time scale and the light propagating time scale of the characteristic length scale $\ell$, respectively, thus $t_\ell^2 / t_g^2 \sim G \varrho / (c^2 k^2)$; $k$ is the wave number introduced as $\Delta = - k^2$. The last condition is our action-at-a-distance assumption; it implies that we keep the action-at-a-distance nature of Newtonian theory in our approximation; in cosmology this condition implies the subhorizon-scale limit \cite{HN-2013-Newtonian}. The reason for having $\gamma^2$ in the last condition will be explained below Eq.\ (\ref{chi-estimate}). The presence of the $\gamma^2$-factor in the action-at-a-distance condition is consistent with the weak gravity condition considering Eqs.\ (\ref{Poisson-ZSG}) and (\ref{Poisson-Psi}) in the regime of ultra-relativistic velocity ($v \sim c$, thus $\gamma \gg 1$).

In the new approximation we {\it ignore} the dimensionless quantities in Eq.\ (\ref{assumptions}) compared with order unity, but consider fully relativistic pressure as well as velocity. On the other hand, the first PN (1PN) approximation takes into account of the first order corrections in the dimensionless quantities of Eq.\ (\ref{assumptions}) as well as first orders in $v^2/c^2$, $p/(\varrho c^2)$, etc. Thus, the two approximations are complementary.

For non-relativistic velocities (slow-motion limit $v^2/c^2 \rightarrow 0$) Eqs.\ (\ref{E-conservation})-(\ref{Poisson-ZSG}) become \cite{HN-2013-pressure}
\bea
   & & \hskip -.5cm
       {d \over dt} \varrho
       + \left( \varrho + {p \over c^2} \right)
       \nabla \cdot {\bf v}
       = {1 \over c^2} {\bf v} \cdot \nabla p,
   \label{E-conservation-NR} \\
   & & \hskip -.5cm
       {d {\bf v} \over dt}
       = \nabla \Phi
       - {1 \over \varrho + p/c^2}
       \left( \nabla p
       + {1 \over c^2} {\bf v} \dot p \right),
   \label{M-conservation-NR} \\
   & & \hskip -.5cm
       \Delta \Phi
       + 4 \pi G \varrho
       = 0.
   \label{Poisson-NR}
\eea
In this case we have $\Psi = \Phi$, see Eqs.\ (\ref{Phi-Psi}) and (\ref{Phi-Psi-2}). Further assuming non-relativistic pressure (thus $c \rightarrow \infty$ limit) we recover the well known Newtonian hydrodynamic equations with gravity \cite{HN-2013-Newtonian}
\bea
   & & {\partial \varrho \over \partial t}
       + \nabla \cdot \left( \varrho {\bf v} \right)
       = 0, \quad
       {\partial {\bf v} \over \partial t}
       + {\bf v} \cdot \nabla {\bf v}
       = \nabla \Phi
       - {1 \over \varrho} \nabla p,
   \nonumber \\
   & & \Delta \Phi
       + 4 \pi G \varrho
       = 0.
\eea

%
%

\vskip .1cm
{\bf 3. Proof:}
Now we derive Eqs.\ (\ref{E-conservation})-(\ref{Poisson-ZSG}) and (\ref{Poisson-Psi}), and show that these are consistent with full Einstein's equations under our assumptions of weak gravity and action-at-a-distance in Eq.\ (\ref{assumptions}).

Our equations are based on the zero-shear gauge. However, as {\it our} zero-shear gauge imposes the temporal gauge (slicing or hypersurface) condition on the $g_{0i}$ setting only its longitudinal part to zero (the three spatial gauge conditions are imposed on the $g_{ij}$ part) \cite{HN-2013-FNL}, the simple form of the metric in Eq.\ (\ref{metric}) demands explanation. For a {\it proper} derivation we need to consider a more general form of the metric written as
\bea
   & & ds^2 = - \left( 1 - {2 \Phi \over c^2} \right) c^2 d t^2
       + \left( 1 + {2 \Psi \over c^2} \right) \delta_{ij} d x^i d x^j
   \nonumber \\
   & & \qquad
       - 2 \chi_i c dt d x^i,
   \label{metric-2}
\eea
where the index of $\chi_i$ is raised and lowered by $\delta_{ij}$ as the metric. We have presented the fully nonlinear and exact perturbation formulation based on the metric in Eq.\ (\ref{metric-2}) extended to the cosmological context \cite{HN-2013-FNL,Noh-2014}. Compared with notations in \cite{HN-2013-FNL,Noh-2014,HN-2013-pressure,HN-2013-Newtonian} we have
\bea
   \alpha \equiv - {\Phi \over c^2}, \quad
       \varphi \equiv {\Psi \over c^2},
   \label{identification}
\eea
with the cosmic scale factor $a \equiv 1$ and the spatial comoving part of Robertson-Walker metric becoming $\gamma_{ij} \equiv \delta_{ij}$. In our derivation we will use the exact Einstein's field equations presented in Sec.\ 3 of \cite{Noh-2014}.

The spatial part of the metric is simple because we have {\it ignored} the transverse-tracefree part of the perturbation (this is a physical assumption), and have {\it imposed} the three spatial gauge conditions without losing generality to fully nonlinear order \cite{Bardeen-1988,HN-2013-FNL}. The zero-shear hypersurface (temporal gauge) condition imposes the longitudinal part of $\chi_i$ to be zero: as we decompose $\chi_i \equiv \chi_{,i} + \chi^{(v)}_i$ with $\chi^{(v)i}_{\;\;\;\;\;\;,i} \equiv 0$, the zero-shear gauge sets $\chi \equiv 0$, thus we still have non-vanishing $\chi_i = \chi^{(v)}_i$ and this should be considered properly.

We consider a fluid energy-momentum tensor {\it without} anisotropic stress
\bea
   T_{ab} = \left( \mu + p \right) u_a u_b + p g_{ab},
\eea
where $\mu \equiv \varrho c^2$ is the energy density. For the fluid four-vector $u_c$ we introduce\footnote{
          Compared with previous notation, our $v_i = \widehat v_i$ in \cite{HN-2013-FNL,Noh-2014,HN-2013-pressure,HN-2013-Newtonian}, see the Appendix D in \cite{HN-2013-FNL}.}
\bea
   u_i \equiv \gamma {v_i \over c}, \quad
        \gamma \equiv - n_c u^c
        = {1 \over \sqrt{ 1 - {v^k v_k
       \over c^2 (1 + 2 \Psi/c^2) }}},
   \label{v_i}
\eea
where the index of $v_i$ is raised and lowered by $\delta_{ij}$ as the metric; $n_a$ is the normal-frame four-vector. In our weak gravity approximation the above fully nonlinear expression of the Lorentz factor becomes the familiar one in Eq.\ (\ref{gamma}).

%
%
Using only the weak gravity condition, the ADM (Arnowitt-Deser-Misner) momentum constraint equation in Eq.\ (3.3) of \cite{Noh-2014} gives
\bea
   & & \kappa_{,i}
       + {3 \over 4} c \Delta \chi^{(v)}_i
       = - {12 \pi G \over c^2} \left(
       \varrho + {p \over c^2} \right)
       \gamma^2 v_i,
   \label{mom-constraint}
\eea
where $\kappa$ is defined as the trace of extrinsic curvature which is the same as the expansion scalar of the normal frame with a negative sign, thus $\kappa \equiv c K^i_i = - \theta^{(n)} \equiv - c n^c_{\;\; ;c}$
\cite{Bardeen-1988,HN-2013-FNL}. Equation (\ref{mom-constraint}) can be decomposed as
\bea
   & & \kappa
       = - {12 \pi G \over c^2} \Delta^{-1} \nabla^i
       \left[ \left(
       \varrho + {p \over c^2} \right)
       \gamma^2 v_i \right],
   \label{mom-constraint-kappa} \\
   & & c \Delta \chi^{(v)}_i
       = - {16 \pi G \over c^2} \bigg\{
       \left(
       \varrho + {p \over c^2} \right)
       \gamma^2 v_i
   \nonumber \\
   & & \qquad
       - \nabla_i \Delta^{-1} \nabla^j
       \left[ \left(
       \varrho + {p \over c^2} \right)
       \gamma^2 v_j \right] \bigg\}.
   \label{mom-constraint-chi}
\eea
From these we can estimate
\bea
   \nabla_i \kappa / c
       \sim \Delta \chi_i^{(v)}
       \sim  \left( \gamma^2 {t_\ell^2 \over t_g^2} \right)
       \Delta v_i/c.
   \label{chi-estimate}
\eea
Using the action-at-a-distance condition in Eq.\ (\ref{assumptions}) we can ignore $\kappa$ and $\chi_i^{(v)}$ compared with $\nabla \cdot {\bf v}$ and $v_i/c$, respectively; these estimates, which follow from the action-at-a-distance assumption in Eq.\ (\ref{assumptions}), are demanded in the following calculation, and this justifies the presence of $\gamma^2$-factor in our relativistic action-at-a-distance condition in Eq.\ (\ref{assumptions}).

%
%
Using Eqs.\ (\ref{assumptions}) and (\ref{chi-estimate}), the covariant momentum conservation in Eq.\ (3.9) of \cite{Noh-2014} gives
\bea
   & & {1 \over \gamma} {d \over dt} \left( \gamma {\bf v} \right)
       = \nabla \Phi
       + {v^2 \over c^2} \nabla \Psi
   \nonumber \\
   & & \qquad
       - {1 \over \varrho + p/c^2}
       \left( {1 \over \gamma^2} \nabla p
       + {\bf v} {1 \over c^2} {d {p} \over dt} \right),
   \label{M-conservation-1}
\eea
with $\gamma = 1/\sqrt{1-v^2/c^2}$.

Notice the presence of post-Newtonian potential $\Psi$ in Eq.\ (\ref{M-conservation-1}) with a $v^2/c^2$-factor. We point out that although the $(v^2/c^2)\nabla\Psi$ term in Eq.\ (\ref{M-conservation-1}) is comparable to the $\nabla\Phi$ term in our approximation, it is negligible compared with the convective term ${\bf v} \cdot \nabla {\bf v}$ in the left-hand-side due to the weak gravity condition. This apparent conflict can be resolved as the following. The $(v^2/c^2)\nabla\Psi$ term is comparable to the $\nabla\Phi$ term either for ultra-relativistic ($v^2 \sim c^2$) or for the relativistic (we keep $v^2/c^2$ order but $v^2/c^2 \ll 1$) velocities; the term is naturally negligible for non-relativistic velocity. In the first (ultra-relativistic) case the $\nabla\Phi$ term itself is negligible (by the weak gravity condition) compared with the convective term, thus the $(v^2/c^2)\nabla\Psi$ term can be ignored as well. In the second (relativistic) case we have the $(v^2/c^2)\nabla\Psi$ much smaller than the $\nabla\Phi$, and the latter is again much smaller than the convective term: thus although we may have to keep the $\nabla\Phi$ term, the $(v^2/c^2)\nabla\Psi$ term can be ignored.

Thus ignoring the $\Psi$ term in Eq.\ (\ref{M-conservation-1}), it can be arranged to give
\bea
   {d \over dt} \ln{\gamma}
       = - {1 \over \varrho + p/c^2}
       {1 \over c^2}
       \left( {d p \over dt}
       - {1 \over \gamma^2} \dot {p} \right),
   \label{M-conservation-3}
\eea
and Eq.\ (\ref{M-conservation}).

Using Eq.\ (\ref{M-conservation-3}), the covariant energy conservation in Eq.\ (3.8) of \cite{Noh-2014} gives Eq.\ (\ref{E-conservation}).

For later use, we present alternative expressions of the conservation equations
\bea
   & & \hskip -.5cm
       {\partial \over \partial t} \left[ \left(
       \varrho + {p \over c^2} \right) \gamma^2 \right]
       + \nabla \cdot \left[ \left(
       \varrho + {p \over c^2} \right)
       \gamma^2 {\bf v} \right]
       = {1 \over c^2} \dot {p},
   \label{E-conservation-2} \\
   & & \hskip -.5cm
       {\partial \over \partial t} \left[
       \left( \varrho + {p \over c^2} \right) \gamma^2 {\bf v} \right]
       + \nabla_j \left[
       \left( \varrho + {p \over c^2} \right) \gamma^2 v^j {\bf v} \right]
   \nonumber \\
   & & \qquad \hskip -.5cm
       = \left( \varrho + {p \over c^2} \right) \gamma^2 \nabla \Phi
       - \nabla p.
   \label{M-conservation-4}
\eea

In the absence of gravity the special relativistic hydrodynamic equations are properly recovered\footnote{
          In the absence of gravity, Eq.\ (\ref{M-conservation}) reduces to Eq.\ (2.10.16) in \cite{Weinberg-1972}, and Eq.\ (\ref{E-conservation-2}) reduces to Eq.\ (2.65) in \cite{Battaner-1996}.
          }.
As far as we are aware, the special relativistic hydrodynamic equations including gravity is unknown in the literature. Our equations can be regarded as the special relativistic hydrodynamic equations in the presence of weak gravity with the action-at-a-distance limit. Our equations, however, are valid in the presence of fully general relativistic pressure and velocity. Whether such an asymmetric (weak gravity and action-at-a-distance limit associated with relativistic pressure and velocity) situation is allowed in Einstein's gravity requires the analysis of full Einstein's equations which we will embark in the following.

%
%
We derive the Poisson's equations and check the consistency of complete Einstein's equations. Using Eqs.\ (\ref{assumptions}) and (\ref{chi-estimate}), the ADM energy-constraint equation and the trace of ADM propagation equation in Eqs.\ (3.2) and (3.4), respectively, of \cite{Noh-2014} give
\bea
   & & \hskip -.8cm
       \Delta \Psi
       + 4 \pi G \varrho
       =
       - 4 \pi G \left(
       \varrho + {p \over c^2} \right)
       \left( \gamma^2 - 1 \right),
   \label{Poisson-1} \\
   & & \hskip -.8cm
       \Delta \Phi
       + 4 \pi G \left( \varrho + 3 {p \over c^2} \right)
       = \dot \kappa
       - 8 \pi G \left(
       \varrho + {p \over c^2} \right)
       \left( \gamma^2 - 1 \right).
   \label{Poisson-2}
\eea
Equation (\ref{Poisson-1}) is the same as Eq.\ (\ref{Poisson-Psi}), but Eq.\ (\ref{Poisson-2}) needs further analysis.
In order to show the consistency of these two equations and to derive Eq.\ (\ref{Poisson-ZSG}), we need the tracefree part of ADM propagation equation in Eq.\ (3.5) of \cite{Noh-2014}. It gives
\bea
   & & \hskip -.8cm
       \left( \nabla^i \nabla_j
       - {1 \over 3} \delta^i_j \Delta \right)
       \left( \Phi - \Psi \right)
   \nonumber \\
   & & \qquad \hskip -.8cm
       - 8 \pi G \left(
       \varrho + {p \over c^2} \right)
       \gamma^2 \left(
       {v^i v_j \over c^2}
       - {1 \over 3} {v^2 \over c^2} \delta^i_j \right)
   \nonumber \\
   & & \qquad \hskip -.8cm
       = - {c \over 2} {\partial \over \partial t}
       \left( \chi^{(v)i}_{\;\;\;\;\;\;,j}
       + \chi^{(v),i}_j \right)
   \nonumber \\
   & & \qquad \hskip -.8cm
       = {8 \pi G \over c^2} \Delta^{-1} \bigg\{
       - \nabla_j \nabla_k \left[ \left( \varrho
       + {p \over c^2} \right) \gamma^2 v^i v^k \right]
   \nonumber \\
   & & \qquad \qquad \hskip -.8cm
       - \nabla^i \nabla_k \left[ \left( \varrho
       + {p \over c^2} \right) \gamma^2 v_j v^k \right]
   \nonumber \\
   & & \qquad \qquad \hskip -.8cm
       + 2 \nabla^i \nabla_j \Delta^{-1}
       \nabla_k \nabla_\ell \left[
       \left( \varrho + {p \over c^2} \right) \gamma^2 v^k v^\ell \right] \bigg\},
   \label{Phi-Psi}
\eea
where we used Eqs.\ (\ref{mom-constraint-chi}) and (\ref{M-conservation-4}). From this we have
\bea
   & & \Delta \left( \Phi - \Psi \right)
       = 12 \pi G \Delta^{-1} \nabla_i \nabla^j
   \nonumber \\
   & & \qquad \times\left\{
       \left( \varrho + {p \over c^2} \right)
       \gamma^2 \left(
       {v^i v_j \over c^2}
       - {1 \over 3} {v^2 \over c^2} \delta^i_j \right)
       \right\}.
   \label{Phi-Psi-2}
\eea
We note that $\chi^{(v)}_i$ has a nontrivial role only in Eq.\ (\ref{Phi-Psi}) but the term has no role in deriving Eq.\ (\ref{Phi-Psi-2}). In this sense the metric can be written as in Eq.\ (\ref{metric}).

Now, from Eqs.\ (\ref{mom-constraint-kappa}) and  (\ref{M-conservation-4}) we can show
\bea
   & & \hskip -.5cm
       \dot \kappa
       = 12 \pi G \Delta^{-1} \nabla_i \nabla_j
       \left[ \left(
       \varrho + {p \over c^2} \right)
       \gamma^2 {v^i v^j \over c^2} \right]
       + 12 \pi G {p \over c^2}
   \nonumber \\
   & & \qquad \hskip -.5cm
       - {12 \pi G \over c^2} \Delta^{-1} \nabla \cdot \left[
       \left( \varrho + {p \over c^2} \right) \gamma^2
       \nabla \Phi \right].
   \label{dot-kappa}
\eea
Notice the consequent {\it cancelation} of the $12 \pi G p/c^2$ term on the left-hand-side of Eq.\ (\ref{Poisson-2}) with the one in $\dot \kappa$. Using Eqs.\ (\ref{Phi-Psi-2}) and (\ref{dot-kappa}), Eqs.\ (\ref{Poisson-1}) and (\ref{Poisson-2}) are consistent with each other, and Eq.\ (\ref{Poisson-2}) can be arranged to the Newtonian Poisson's equation in Eq.\ (\ref{Poisson-ZSG}).

We have one remaining equation to be checked which is the definition of $\kappa$ (the trace of extrinsic curvature) equation in Eq.\ (3.1) of \cite{Noh-2014}. It gives
\bea
   \kappa = - 3 {\dot \Psi \over c^2}.
   \label{kappa-def}
\eea
Using Eqs.\ (\ref{mom-constraint-kappa}), (\ref{E-conservation-2}) and (\ref{Poisson-1}) we can show that Eq.\ (\ref{kappa-def}) is identically satisfied. This {\it completes} the consistency check of full Einstein's equations.

In our calculation, except for the Eq.\ (\ref{Phi-Psi}), the $\chi^{(v)}_i$ terms appearing in various forms are all negligible due to the estimate in Eq.\ (\ref{chi-estimate}): as a consequence the metric can be written as in Eq.\ (\ref{metric}), i.e., the $\chi_i$ can be ignored due to the action-at-a-distance assumption in addition to the zero-shear gauge condition. This completes our {\it proof} of Eqs.\ (\ref{E-conservation})-(\ref{Poisson-ZSG}), (\ref{Poisson-Psi}) with the metric in Eq.\ (\ref{metric}) under conditions in Eq.\ (\ref{assumptions}).


%
%

\vskip .1cm
{\bf 4. Discussion:}
Our new hydrodynamic approximation method proposed in Eqs.\ (\ref{E-conservation})-(\ref{Poisson-ZSG}) is derived consistently from Einstein's gravity and is complementary to the 1PN approximation \cite{Chandrasekhar-1965,Weinberg-1972,HNP-2008-PN}. Here we consider fully relativistic pressure and velocity while taking weak gravity and action-at-a-distance limit, whereas in the 1PN approximation we consider only 1PN orders in both the pressure and the velocity, but consider both the gravitational potential and the assumption on the action-at-a-distance to the 1PN orders as well, thus giving time-delayed propagation of gravity correction to the action-at-a-distance nature of the Newtonian limit (0PN), see sections 6 and 7 in \cite{HNP-2008-PN}.

Our equations are valid in the zero-shear gauge. As our gauge conditions (spatial and temporal) completely fix the gauge mode there remains no remnant gauge mode, and each variable has a corresponding gauge-invariant expression to fully nonlinear order, see \cite{Bardeen-1988,HN-2013-FNL}. Our equations are consistently derived from Einstein's gravity, with assumptions in Eq.\ (\ref{assumptions}). The 1PN hydrodynamic equations in \cite{Chandrasekhar-1965} and section 9.8 of \cite{Weinberg-1972} are presented in the uniform-expansion gauge and the harmonic gauge, respectively. Extension to the general gauge condition is made in \cite{HNP-2008-PN}. We have checked that the 1PN equations of \cite{HNP-2008-PN} in the zero-shear gauge are consistent with Eqs.\ (\ref{E-conservation})-(\ref{Poisson-ZSG}) and (\ref{Poisson-Psi}) in the overlapping regimes (i.e., 0PN for the weak gravity and the action-at-a-distance, and 1PN for the pressure and velocity) of the two approximations. We are currently studying the same problem in other gauge conditions where the form of equations varies for both our approximation and the 1PN approximation \cite{HN-2016}.

Applications to astrophysical situations (requiring hydrodynamic numerical simulation) with relativistic pressure medium and/or relativistic velocity might be easier based on our equations, as long as the weak gravity and action-at-a-distance assumptions are met. Our newly proposed equations basically have the Newtonian hydrodynamic structure which is much simpler than the fully general relativistic system.

%
%
\vskip .1cm
\noindent {\bf Acknowledgments}
J.H.\ was supported by Basic Science Research Program through the National Research Foundation (NRF) of Korea funded by the Ministry of Science, ICT and future Planning (No. 2013R1A2A2A01068519). He also wish to thank hospitality during his visit UFES in Vit\'oria and acknowledge support by CNPq (Brazil).
H.N.\ was supported by National Research Foundation of Korea funded by the Korean Government (No.\ 2015R1A2A2A01002791).
O.F.P.\, W.Z.\ and J.F.\ thank CNPq and FAPES (Brazil) for partial financial support.

%
%



\begin{thebibliography}{99}
\bibitem{numerical-relativity}
         J.R. Wilson, G.J. Mathews, {\it Relativistic numerical hydrodynamics} (Cambridg Univ. Press., Cambridge, 2003);
         M. Alcubierre, {\it Introduction to 3+1 numerical relativity}, (Oxford Univ. Press., New York, 2008);
         C. Bona, C. Palenzuela-Luque, C. Bona-Casas, {\it Elements of Numerical Relativity and Relativistic Hydrodynamics: From Einstein's Equations to Astrophysical Simulations} (Springer, 2009);
         T.W. Baumgarte, S.L. Shapiro, {\it Numerical Relativity: Solving Einstein's Equations on the Computer} (Cambridge Univ. Press, 2010);
         E. Gourgoulhon, {\it 3+1 Formalism in General Relativity: Bases of Numerical Relativity} (Springer, 2012);
         L. Rezzolla, O.Zanotti, {\it Relativistic Hydrodynamics} (Oxford Univ. Press, 2013)
\bibitem{Chandrasekhar-1965}
         S. Chandrasekhar, Astrophys. J. {142} (1965) 1488.
\bibitem{Weinberg-1972}
         S. Weinberg, {\it Gravitation and Cosmology} (Wiley, New York, 1972).
\bibitem{PN}
         V. A. Fock, {\it The Theory of Space, Time and Gravitation} (Oxford, Pergamon, 1964);
         L. Blanchet, T. Damour, G. Sch\"afer, Mon. Not. R. Astron. Soc. 242 (1990), 289;
         M, Shibata, H. Asada, Prog. Theor. Phys. 94 (1995) 11;
         H. Asada, T. Futamase, Prog. Theor. Phys. Suppl. 128 (1997) 123;
         M. Takada, T. Futamase, Mon. Not. R. Astron. Soc. 306 (1999) 64.
\bibitem{HNP-2008-PN}
         J. Hwang, H. Noh, D. Puetzfeld, JCAP {03} (2008) 010.
\bibitem{HN-2013-Newtonian}
         J. Hwang, H. Noh, JCAP {04} (2013) 035.
\bibitem{HN-2013-pressure}
         J. Hwang, H. Noh, JCAP {10} (2013) 054.
\bibitem{HN-2013-FNL}
         J. Hwang, H. Noh, MNRAS {433} (2013) 3472.
\bibitem{Noh-2014}
         H. Noh, JCAP {07} (2014) 037.
\bibitem{Bardeen-1988}
         J.M. Bardeen, in {\it Particle Physics and Cosmology}, edited by
         L. Fang and A. Zee (Gordon and Breach, London, 1988), p1.
\bibitem{Battaner-1996}
         E. Battaner, {\it Astrophysical Fluid Dynamics} (Cambridge University Press, Cambridge, England, 1996).
\bibitem{HN-2016}
         J. Hwang, H. Noh, in preparation (2016).
\end{thebibliography}
\end{document}